\documentclass[prd,reprint,letterpaper,superscriptaddress,amsfonts,amsmath,amssymb,nofootinbib]{revtex4-2}

\usepackage{braket}

\usepackage{graphicx}
\usepackage{color}
\usepackage[normalem]{ulem}
\usepackage{MnSymbol}
\usepackage{enumerate}
\usepackage{bbold}
\usepackage{subfigure}
\usepackage{subfiles}
\usepackage{units}
\usepackage[usenames,dvipsnames]{xcolor}
\usepackage[colorlinks=true,linkcolor=MidnightBlue,citecolor=blue,filecolor=green,urlcolor=Violet]{hyperref}
\usepackage[dvipsnames]{xcolor}
\usepackage{paralist}
\usepackage{psfrag}
\usepackage{setspace,tabularx,fancyhdr}
\usepackage{mathtools}
\usepackage[top=2cm,bottom=2cm,left=2cm,right=2cm]{geometry}
\usepackage{enumitem}
\setlist[itemize]{leftmargin=*}
\usepackage{wrapfig}
\usepackage{type1cm}
\usepackage{yfonts}
\usepackage{wasysym}

\def\beq{\begin{equation}}
\def\eeq{\end{equation}}
\def\bsp{\begin{split}}
\def\esp{\end{split}}
\def\bea{\begin{eqnarray}}
\def\eea{\end{eqnarray}}

\definecolor{myyellow}{rgb}{0.94, 0.86, 0.51}

\definecolor{mygreen}{rgb}{0.2, 0.8, 0.2}

\definecolor{mypink}{rgb}{0.99, 0, 0.99}

\definecolor{mypurple}{rgb}{0.75, 0, 0.75}

\definecolor{cadmiumorange}{rgb}{0.93, 0.53, 0.18}

\newcommand{\IGNORE}[1]{}

\graphicspath{{Images/}}

\newcommand{\MyCircleArrowRight}{\rightturn}

\newcommand{\MyCircleArrowLeft}{\leftturn}

\begin{document}

\title{Are the Circular Polarizations of Observed Gravitational-Waves
Even-Handed?}

\author{Elena Emtsova}\email{emtsova.e@gmail.com}
\affiliation{Department of Physics, Bar Ilan University, Ramat Gan 5290002, Israel}
\affiliation{Physics Department, Ariel University, Ariel 40700, Israel}

\author{Ofek Birnholtz}
\affiliation{Department of Physics, Bar Ilan University, Ramat Gan 5290002, Israel}

\begin{abstract}
We study whether gravitational waves (GWs) from binary black hole (BBH) mergers show a difference between right- and left-handed circular polarizations. Such a difference could point to a violation of parity symmetry in gravity. We analyze publicly available data from the collaboration of the LIGO-Virgo-KAGRA (LVK) detectors, focusing on events with accurate sky localization using all three detectors. This allows us to separate the circular polarization modes and measure their amplitudes.
Out of 15 well-localized events, 7 show a dominant right-handed mode and 8 a left-handed one. This small difference is consistent with statistical fluctuations. A Bayesian analysis shows that the probability \( p_R \) of a right-handed signal follows a Beta(8,9) distribution, with no strong evidence for asymmetry.
We also estimate how many future events would be needed to detect a small asymmetry like $p_R = 7 / 15$ with high confidence. To reach 2$\sigma$ or 3$\sigma$ significance, about 900 to 2000 well-localized events would be required. These numbers could be reached in upcoming observation runs. Our results support parity symmetry for now, but future data may allow us to test it more precisely.
\end{abstract}


\maketitle

\section{Introduction}
\label{intro}

The observations of gravitational waves (GWs) \cite{LIGO2016,LIGOScientific:2016vbw,LIGOScientific:2016sjg,O1:BBH,GWTC1,GWTC2,GWTC3} by the network of gravitational wave detectors \cite{IGWN} LIGO \cite{aLIGO}, Virgo \cite{VIRGO} and KAGRA \cite{KAGRA,KAGRA2} - collectively LVK - now allow for the first time to search for possible circular polarization symmetry breaking in General Relativity, hinting at deviations from it or corrections to it.
The concept of symmetry breaking plays a pivotal role in many areas of physics, from the weak interactions in the Standard Model to cosmological models explaining matter-antimatter asymmetry.

In the Standard Model, weak interactions violate parity symmetry, meaning they distinguish between left- and right-handed chiral states. This was first discovered in 1956 by Chien-Shiung Wu’s experiment, which demonstrated that beta decay is asymmetric under mirror reflection \cite{Wu1957}.  A gravitational analogue of the Wu experiment—where the sign of the net circular polarization (Stokes V) is tied to the projection of the recoil (kick) onto the final spin—has been formulated explicitly \cite{Ng:2023jjt,Leong:2025raf,CalderonBustillo:2024akj} for related propagation-based parity tests. Similar symmetry-breaking mechanisms might manifest in gravitational interactions, particularly through parity-violating extensions to General Relativity, such as \newline Chern-Simons gravity \cite{Alexander2009} or ghost-free massive gravity \cite{deRham2010}.

    Recent studies have explored various aspects of gravitational wave (GW) data analysis and waveform modeling. The LVK collaboration has provided comprehensive catalogs of GW events, offering posterior samples and parameter estimations for numerous binary mergers \cite{O1:BBH,GWTC1,GWTC2,GWTC3}. These datasets have been instrumental in advancing our understanding of compact binary coalescences.

A growing body of work has explored possible deviations from General Relativity through polarization and aniso\-tropy studies of gravitational waves. Several studies have tested the isotropy of gravitational-wave sources and their properties: \cite{Isi:2023dlk} examined spatial distribution and angular momentum alignment of binary black holes using dipolar models and found consistency with isotropy in O3 data, while \cite{Vitale:2022pmu} confirmed the randomness of orbital orientations in \newline GWTC-3, and \cite{Essick:2022slj} placed the most stringent constraints to date on anisotropies in the sky distribution of mergers, though without a conclusive preference for isotropy. Theoretical motivations for such studies include parity-violating extensions of GR, reviewed in \cite{Qiao:2022mln}, where left-right polarization asymmetries in gravitational waves could arise from new physics, such as Chern-Simons gravity. These ideas are tested observationally in \cite{Okounkova:2021xjv,Ng:2023jjt}, where amplitude birefringence is constrained in GWTC-2 and GWTC-3 data respectively, and in \cite{Haegel:2022ymk,ONeal-Ault:2023lxi}, which use the Standard-Model Extension framework to search for Lorentz- and CPT-violating birefringent and dispersive effects in GW propagation. See also \cite{Yamada:2020zvt} for a parametrized test of parity-violating gravity with the events of GWTC-1. In addition, \cite{CalderonBustillo:2024akj}  measured a net circular polarization observable across 47 black-hole mergers and find that the average across events is consistent with zero; they identify only one event (GW200129) with strong evidence for asymmetry, while the rest show only mild evidence. From a quantum field theory perspective, \cite{Agullo:2017pyg} shows that curved spacetime can itself induce polarization rotation in electromagnetic radiation via quantum anomalies. Finally, a multimessenger strategy introduced by \cite{Lagos:2024boe} leverages electromagnetic observations of binary neutron star mergers to constrain polarization birefringence during gravitational wave propagation, offering a promising future direction with next-generation detectors.

Gravitational waves are generated by accelerating masses with a changing quadrupole moment, such as merging black holes, neutron stars, or asymmetric supernova explosions. Since GWs are transverse traceless tensor waves \cite{Maggiore:2007ulw}, they have two fundamental polarization states. These can be expressed in either the \textbf{linear polarization basis} (\( h_+, h_\times \)), commonly used in ground-based detectors like LIGO, or the \textbf{circular polarization basis} (\( h_{\MyCircleArrowLeft}, h_{\MyCircleArrowRight} \)), which is more natural for describing waves emitted by rotating systems like binary inspirals. The mathematical definitions of these polarization bases, and their relation, are given in Section~\ref{sec:GWbasics}.

If all gravitational interactions respect parity symmetry, and if there is no parity preference in the distribution of positions and angular momentum of stellar mass black holes in the universe, then there should be no preferred handedness in observed LVK signals. We note that while in \cite{Isi:2023dlk} the directional isotropy of the detected GW mergers is shown, we cannot be sure that such an isotropy will be conserved in larger scales. Recent studies have revealed a statistically significant asymmetry in the spin directions of spiral galaxies across the sky. Analysis of datasets from SDSS, HSC-SSP, and JWST indicates a dipole pattern in the distribution of galaxy handedness, suggesting a preferred cosmological axis. \cite{Shamir2012,Shamir2024HSC} report that right- and left-handed spiral galaxies are unevenly distributed, with the asymmetry strengthening at higher redshifts. These results challenge the cosmological principle of isotropy and suggest either observational biases or new physics.
Also, several theories predict parity-violating modifications, which would introduce an imbalance between $h_{\MyCircleArrowRight}$ and $h_{\MyCircleArrowLeft}$. Such an effect could be caused by interactions with axion-like fields \cite{Adshead2013} or chiral gravitational waves in the early universe \cite{Lue1999}.

Observationally, testing this effect requires high-precision measurements of gravitational wave signals from binary black hole (BBH) mergers. The LVK network provides an excellent opportunity to probe such parity-violating effects. The presence of circular polarization asymmetry can be detected by comparing the statistical distributions of polarization modes over many BBH events \cite{Romano2017}. A significant detection of circular polarization asymmetry would have profound implications for fundamental physics, potentially pointing to new interactions in gravity that go beyond General Relativity. 

Future observing runs of the LVK collaboration, with increasing sensitivity and event rates, will enable more stringent tests of this hypothesis. The next-generation detectors, such as the Einstein Telescope and Cosmic Explorer, will allow the study of gravitational waves with unprecedented precision \cite{Punturo2010}\cite{Reitze2019}. If deviations from parity symmetry are found, it could revolutionize our understanding of gravity and hint at deeper connections between gravity and quantum field theory.

In the remainder of this article, we present the structure and findings of our analysis in detail. 
In Section~\ref{sec:Observed}, we summarize the observed LVK events and define the subset of 
three-detector events used in this study. 
In Section~\ref{sec:GWbasics}, we review the relevant gravitational-wave polarization formalism. 
In Section~\ref{sec:Metho}, we describe the methodology for reconstructing the circular polarizations 
%modes
and computing the amplitude ratios. 
In Section~\ref{sec:Results}, we present the results of the analysis, including handedness classification 
and the statistical analysis of the amplitude ratios.
In Section~\ref{sec:Bayes}, we provide a Bayesian estimate of the underlying polarization asymmetry 
and a forecast for future observing runs. 
Finally, in Section~\ref{sec:Concl}, we summarize our conclusions and outline future work.

\section{Observed LVK Events}\label{sec:Observed}

The LVK has detected a total of 90 confident mergers across the first three observing runs (O1, O2, and O3) \cite{4OGC}. The number of detections has increased significantly due to improvements in detector sensitivity and observation time. In O1 (2015-2016), 3 BBH mergers were detected, including the historic first-ever gravitational wave event, GW150914 \cite{LIGO2016}. In O2 (2016-2017), 8 new mergers were identified, including the first neutron star merger GW170817 \cite{GWTC1} \footnote{We exclude GW170817 from the current analysis, anticipating an all-BNS analysis when enough accumulate to provide meaningful statistics - material interactions might differ from the BBH population}. In O3 (2019-2020), the number of detected events significantly increased , by 79 \cite{GWTC3,4OGC}.

\paragraph{Three-Detector (HLV) Events.}

The number of BBH mergers detected simultaneously by all three detectors—LIGO Hanford (H), LIGO Livingston (L), and Virgo (V)—increased notably as Virgo became fully operational.
In O1, no three-detector (HLV) events were observed since Virgo was not operational. In O2, five events were detected with all three detectors (including GW170817). \cite{GWTC1}.
In O3, this number increased to 48 events detected when all 3 detectors were working \cite{GWTC2,GWTC3,4OGC}.

\paragraph{Future Expectations for Detections.}

With increased detector sensitivity and additional observatories joining, the detection rate of mergers is expected to rise dramatically.
The estimated BBH merger rate at redshift \( z = 0 \) is 16.5 to 18.5 Gpc\(^{-3}\) yr\(^{-1}\), increasing to 24.9 to 30.5 Gpc\(^{-3}\) yr\(^{-1}\) at \( z = 0.2 \) \cite{4OGC}. In the ongoing O4 run, with improved sensitivity and full inclusion of KAGRA, detection rates are projected to exceed one  event per week \cite{LVKfuture}.
The fraction of HLV detections is expected to increase by at least 50\%, with future networks including LIGO India allowing for even better sky localization \cite{GWTC3}.

The steady increase in BBH detections from O1 to O3 demonstrates the growing capabilities of the LVK network.
The transition from a few events per year to multiple detections per week in O4 and beyond will enable deeper studies of black hole populations and tests of fundamental physics.

\section{GW basics}\label{sec:GWbasics}%\setcounter{equation}{0}
We will give a brief introduction according to \cite{Maggiore:2007ulw,Isi:2022mbx}. 

\subsection{Linear Basis in General Relativity}

In General Relativity, gravitational waves (GWs) are characterized by two independent polarization states.
The local effect of GWs can be described by the strain tensor $h_{ij}$, which represents the transverse-traceless (TT) part of the metric perturbation:
\begin{equation}
    h_{ij} = \left( \begin{array}{cccc}
      h_+ & h_\times & 0 \\ h_\times & -h_+ & 0 \\ 0 & 0 & 0     \end{array} \right)
    .
\end{equation}
The two fundamental polarization states, known as the \textit{plus} ($+$) and \textit{cross} ($\times$) modes, are defined in terms of the polarization basis tensors in a Cartesian coordinate system where the $z$-axis is chosen along the wave propagation direction:
\begin{equation}
    e^+_{ij} = x_i x_j - y_i y_j, ~~~
    e^\times_{ij} = x_i y_j + y_i x_j,
\end{equation}where $x$ and $y$ are arbitrary orthonormal vectors that,
with $z$, form a right-handed Cartesian basis.
Thus, the strain tensor can be rewritten in terms of these polarization basis tensors as:
\begin{equation}
    h_{ij} = h_+ e^+_{ij} + h_\times e^\times_{ij}.
\end{equation}

Since gravitational waves interact with detectors via their strain, the observed signal $h(t)$ at an interferometric detector such as LIGO or Virgo is given by the projection of $h_{ij}$ onto the detector tensor $D^{ij}$:
\begin{equation}\label{strainantenna}
    h(t) = D^{ij} h_{ij} = F_+ h_+ + F_\times h_\times,
\end{equation}
where $F_+$ and $F_\times$ are known as the \textit{antenna response functions} of the detector.
These functions depend on the relative orientation of the detector arms and the incoming wave's propagation direction. For a detector with arms aligned along unit vectors $X^i$ and $Y^i$, the detector tensor is given by:
\begin{equation}
    D^{ij} = \frac{1}{2} (X^i X^j - Y^i Y^j).
\end{equation}
The response functions can be computed as:
\begin{equation}
    F_+ = D^{ij} e^+_{ij}, ~~~
    F_\times = D^{ij} e^\times_{ij}.
\end{equation}
These equations encapsulate how an interferometric detector measures a linear combination of the plus and cross polarization components of an incoming gravitational wave. The ability to measure both polarizations is crucial for extracting astrophysical information about the source, such as its inclination angle and distance.

A single gravitational-wave (GW) detector measures a linear combination of the two polarization modes, \( h_+ \) and \( h_\times \), weighted by the antenna response functions \( F_+ \) and \( F_\times \), as described in (\ref{strainantenna}).
Since this provides only one equation for two unknowns, a single detector cannot fully separate the polarizations. With two detectors, such as LIGO Hanford (H1) and LIGO Livingston (L1), we obtain two independent measurements,  
\[
h_{\text{H1}} = F_+^{H1} h_+ + F_\times^{H1} h_\times, \quad h_{\text{L1}} = F_+^{L1} h_+ + F_\times^{L1} h_\times,
\]
which allow for partial polarization determination if the response functions differ. The key factors enabling this are the slight misalignment between H1 and L1 (a 90° rotation) and the time delay between their detections, which provides additional information about the wave's propagation direction. However, due to their similar geometries, and non-certain sky location the detectors do not provide fully independent constraints, resulting in only partial separation of \( h_+ \) and \( h_\times \). The addition of a third detector, such as Virgo (or KAGRA), significantly improves polarization reconstruction by introducing an additional independent equation, allowing for full determination of the polarization states and improved source localization through triangulation
\cite{Isi:2022mbx}.

\subsection{Circular Basis in General Relativity}

Instead of the linear plus and cross polarizations, we can define an equivalent circular polarization basis.
The right-handed ($h_{\MyCircleArrowRight}$) and left-handed ($h_{\MyCircleArrowLeft}$) circular polarizations are defined in terms of the linear polarizations as:
%\FIX{this discussion seems to repeat a lot of the stuff in the introduction, but with 1. a different convention (R/L vs. $\MyCircleArrowRight$/$\MyCircleArrowLeft$)  2. with different definitions of $1/\sqrt{2}$ vs. $1/2$ }
\begin{equation}
e^{\MyCircleArrowRight / \MyCircleArrowLeft}_{ij} = e^+_{ij} \mp i e^\times_{ij}.
\end{equation}
In this basis, the strain tensor can be expressed as:
\begin{equation}
    h_{ij} = h_{\MyCircleArrowRight} e^R_{ij} + h_{\MyCircleArrowLeft} e^L_{ij},
\end{equation}
where the complex polarization amplitudes are given by:
\begin{equation}\label{polarisations}
    \begin{array}{cccc}
            h_{\MyCircleArrowLeft} &=& h_+ + i h_\times,
    \quad ~~
    h_{\MyCircleArrowRight} = h_+ - i h_\times
    ; \\
    h_+ &=& \frac{h_{\MyCircleArrowLeft} + h_{\MyCircleArrowRight}}{2},
    \quad 
    h_\times = \frac{h_{\MyCircleArrowLeft} - h_{\MyCircleArrowRight}}{2i}.
\end{array}
\end{equation}

The right-handed mode $h_{\MyCircleArrowRight}$ and the left-handed mode $h_{\MyCircleArrowLeft}$ correspond to waves that rotate counterclockwise and clockwise, respectively, when observed head-on.
The circular polarization basis is particularly useful in Fourier space, where these modes transform under rotations as:
\begin{equation}
    h_{\MyCircleArrowRight}' = h_{\MyCircleArrowRight} e^{-i2\psi}, ~~~ h_{\MyCircleArrowLeft}' = h_{\MyCircleArrowLeft} e^{i2\psi}.
\end{equation}
This transformation arises because gravitational waves are spin-2 fields, meaning they acquire a phase shift of $2\psi$ under a coordinate rotation by $\psi$.
The right-handed polarization $h_{\MyCircleArrowRight}$ rotates clockwise, while the left-handed polarization $h_{\MyCircleArrowLeft}$ rotates counterclockwise.
Since circularly polarized waves are eigenstates of rotation, this property makes them particularly useful in cosmological and astrophysical analyses, where parity-violating effects or interactions with chiral fields may modify their evolution.

This property makes circular polarizations advantageous for analyzing gravitational wave signals from astrophysical sources such as compact binary mergers, where the emission tends to be dominated by a single handedness depending on the inclination of the system.

\subsubsection{Physical Interpretation of Circular Polarization}

A purely right-handed circularly polarized wave (R-mode) follows a counterclockwise rotation pattern when observed along the propagation direction, whereas a left-handed wave (L-mode) follows a clockwise rotation.
This is best understood by analyzing the phase evolution of the components:
\begin{equation}
    h_{\MyCircleArrowRight}(t) = A_{\MyCircleArrowRight} e^{-i \omega t}, \quad h_{\MyCircleArrowLeft}(t) = A_{\MyCircleArrowLeft} e^{i \omega t},
\end{equation}
where $A_{\MyCircleArrowRight}$ and $A_{\MyCircleArrowLeft}$ are the amplitude factors of the right- and left-handed modes, respectively.
A general gravitational wave signal can be described as a superposition of these circularly polarized components.

\subsubsection{Application to Astrophysical Sources}

Gravitational waves from astrophysical sources such as binary mergers often exhibit dominant circular polarization at specific inclination angles.
For instance, a face-on or face-off binary system (where the orbital plane is perpendicular to the line of sight) produces purely circularly polarized waves, while an edge-on system produces linearly polarized waves. The inclination angle $\iota$ affects the relative strengths of the $h_{\MyCircleArrowRight}$ and $h_{\MyCircleArrowLeft}$ components, leading to an observed elliptically polarized signal in general cases.

By decomposing the wave into circular components, it is possible to extract valuable information about the binary system's orientation, spin interactions, and potential deviations from General Relativity.

\section{Methodology}\label{sec:Metho}
\subsection{Precise Sky Localization with Three Detectors}

In gravitational wave (GW) detection, accurately determining the source location on the sky relies on measuring the time delays between signal arrivals at multiple detectors.
When a gravitational-wave event is detected by two observatories, the difference in arrival times constrains the source to an annular region on the sky: the set of points consistent with the observed time delay forms a narrow ring. However, the detectors have direction-dependent sensitivity patterns, described by their antenna response functions, which vary across the sky. These antenna patterns modulate the effective probability distribution of the source's location, favoring regions where the detector network is more sensitive. As a result, the initially annular localization is distorted into an elongated, curved structure — a "banana"-shaped region — reflecting both the timing constraint and the varying sensitivity of the detectors to different sky positions.
However, with three or more detectors, the precise sky coordinates can be determined by triangulation only.

Using three detectors such as LIGO-Hanford (H), LIGO-Livingston (L), and Virgo (V), the time delays between wave arrivals provide geometric constraints on the sky position.
The arrival-time delays between pairs of detectors, such as Hanford and Livingston ($\Delta t_{LH}$), and Livingston and Virgo ($\Delta t_{LV}$), enable the determination of the source's latitude ($\theta_{lat}$) and longitude ($\phi_{long}$) via the equations:
\bea
    \Delta t_{LH} &=& \frac{\hat{k}(\theta_{lat}, \phi_{long}) \cdot \mathbf{r}_{LH}}{c},
\\
\Delta t_{LV} &=& \frac{\hat{k}(\theta_{lat}, \phi_{long}) \cdot \mathbf{r}_{LV}}{c},
\eea
where:
- $\hat{k}(\theta_{lat}, \phi_{long})$ is the unit vector in the GW propagation direction,
- $\mathbf{r}_{LH}$ and $\mathbf{r}_{LV}$ are the position vectors between the detectors,
- $c$ is the speed of light \cite{Hilborn:2018rio}.

These two equations provide two independent constraints to solve for the unknowns ($\theta_{lat}, \phi_{long}$), enabling precise localization.
The accuracy depends mainly on the timing precision of the detectors, which typically have arrival-time delay uncertainties on the order of milliseconds.
This translates into sky localization uncertainties of a few degrees \cite{Hilborn:2018rio}.

Using this triangulation method, the LIGO-Virgo collaboration localized the GW170814 event to a 90\% credible region of just 60 deg², significantly smaller than the regions obtained using only two detectors \cite{Hilborn:2018rio}.

Thus, the use of three detectors is crucial for resolving sky position ambiguities, facilitating accurate astrophysical follow-ups and polarization studies.

\subsection{GW Polarizations Depending on Sky Location}

The polarization of gravitational waves is influenced by the source's sky location. The signal observed at a detector can be expressed as:
\bea
    h_D(t) &=& h_1^D(t) \cos 2\Phi(t) + h_2^D(t) \sin 2\Phi(t)
\\
    h_i^D(t) &=& a_i(t) f_i(\theta) \hat{F}_i(\omega, \delta)
\eea
where:
- $D$ represents the detector (H for Hanford, L for Livingston, V for Virgo),
- $a_i(t)$ is a slowly varying amplitude dependent on the source's distance and properties,
- $\Phi(t)$ is the time-dependent orbital phase,
- $f_i(\theta)$ captures the dependence on the orbital inclination angle,
- $\hat{F}_i(\omega, \delta)$ represents the antenna pattern functions, describing the dependence on wave propagation direction, polarization angle, and detector geometry.

\subsubsection{Dependence of $h_+$ and $h_\times$ on Sky Location}

The gravitational wave polarizations, plus ($h_+$) and cross ($h_\times$), depend strongly on the source's sky position.
Specifically:
- The antenna pattern functions $\hat{F}_i(\omega, \delta)$ modulate the observed signals at each detector based on the source's relative position to Earth.
- The observed amplitudes of $h_+$ and $h_\times$ vary with the inclination angle $\theta_i$ and polarization angle $\psi$.
- When $\theta_i \approx 0$ (face-on binary orbit), both polarization components are nearly equal, leading to circular polarization.
- When $\theta_i \to \pi/2$ (edge-on binary orbit), the cross polarization $h_\times$ diminishes, leaving primarily $h_+$, and the polarization angle dependence becomes stronger.

The antenna pattern functions for $h_+$ and $h_\times$ describe the detector response to each polarization mode based on the wave's propagation direction and the detector's orientation:
\bea
    F_+(\hat{w}, \hat{d}) &=& \frac{1}{2} \left[ (\hat{e}_1 \cdot \hat{d}_a)^2 - (\hat{e}_1 \cdot \hat{d}_b)^2 - (\hat{e}_2 \cdot \hat{d}_a)^2 + (\hat{e}_2 \cdot \hat{d}_b)^2 \right], \nonumber
\\
    F_\times(\hat{w}, \hat{d}) &=& (\hat{e}_1 \cdot \hat{d}_a)(\hat{e}_2 \cdot \hat{d}_a) - (\hat{e}_1 \cdot \hat{d}_b)(\hat{e}_2 \cdot \hat{d}_b)
\eea

where:
- $\hat{e}_a$ and $\hat{e}_b$ are the polarization unit vectors associated with the gravitational wave,
- $\hat{d}_a$ and $\hat{d}_b$ define the interferometer arm directions,
- $\hat{w} = \{\hat{e}_1, \hat{e}_2, k\}$ represents the wave propagation frame.

These functions encode how the gravitational wave interacts with the detector geometry. Their dependence on the source's sky location and polarization angle is crucial for precise wave characterization.
\subsection{Calculation Method of Circular Polarizations and Amplitude Ratios}\label{subsec:calcpol}

Our analysis of circular polarization asymmetry in gravitational-wave events leverages the precise sky localization provided by simultaneous detections across three gravitational-wave observatories: LIGO Hanford, LIGO Livingston, and Virgo. Events with clear localization allow robust extraction of polarization information.  So, we restrict ourselves by such events only.

We start by reconstructing the gravitational waveforms in the time domain for each polarization state—the plus ($h_+$) and cross ($h_\times$) modes—based on posterior samples provided by the LVK collaboration. These waveforms are complexified into analytic signals via the Hilbert transform, allowing for straightforward separation of amplitude and phase components.

The right-handed ($h_{\MyCircleArrowRight}$) and left-handed ($h_{\MyCircleArrowLeft}$) circular polarization states are defined from the linear polarization states through (\ref{polarisations}).
The instantaneous amplitude of each polarization component is then obtained from these complex waveforms:
\begin{equation}
    A_{\MyCircleArrowRight}(t) = |h_{\MyCircleArrowRight}(t)|, \qquad
    A_{\MyCircleArrowLeft}(t) = |h_{\MyCircleArrowLeft}(t)|.
\end{equation}
For each gravitational-wave event, we calculate the wei\-ghted mean amplitude and the associated variances across the samples for both polarization states using likelihood-based weights derived from posterior probabilities:
\bea   \bar{A}_{\MyCircleArrowRight/\MyCircleArrowLeft} &=& \frac{\sum_i w_i A_{\MyCircleArrowRight/\MyCircleArrowLeft,i}}{\sum_i w_i},
\\    \sigma^2_{\MyCircleArrowRight/\MyCircleArrowLeft} &=& \frac{\sum_i w_i A^2_{\MyCircleArrowRight/\MyCircleArrowLeft,i}}{\sum_i w_i} - \bar{A}_{\MyCircleArrowRight/\MyCircleArrowLeft}^2,
\eea
where $w_i$ are the likelihood-based weights of samples.

Subsequently, we compute the amplitude ratio of right- to left-handed polarizations ($A_{\MyCircleArrowRight}/A_{\MyCircleArrowLeft}$) as an indicator of circular polarization dominance:
\begin{equation}
    \frac{A_{\MyCircleArrowRight}}{A_{\MyCircleArrowLeft}} \pm \delta\left(\frac{A_{\MyCircleArrowRight}}{A_{\MyCircleArrowLeft}}\right) = 
    \frac{\bar{A}_{\MyCircleArrowRight}}{\bar{A}_{\MyCircleArrowLeft}}
    \pm
    \frac{\bar{A}_{\MyCircleArrowRight}}{\bar{A}_{\MyCircleArrowLeft}}
    \sqrt{
        \left(
            \frac{\sigma_{\MyCircleArrowRight}}{\bar{A}_{\MyCircleArrowRight}}
        \right)^2
        +
        \left(
            \frac{\sigma_{\MyCircleArrowLeft}}{\bar{A}_{\MyCircleArrowLeft}}
        \right)^2
    }.
\end{equation}
Finally, a time-averaged amplitude ratio across the relevant pre-merger interval is calculated as a weighted sum over discrete time grid points. Let \( t \) index the time samples (assuming a uniform time grid). Then,
\begin{equation}\label{averageratio}
    \bar{R} = \frac{\sum_{t} \left( R_t / \sigma_{R_t}^2 \right)}{\sum_{t} \left( 1 / \sigma_{R_t}^2 \right)},
\end{equation}
where \( R_t \) is the amplitude ratio at time \( t \), and \( \sigma_{R_t} \) is its corresponding uncertainty. The uncertainty of this time-averaged ratio is given by
\begin{equation}
    \sigma_{\bar{R}} = \sqrt{ \frac{1}{\sum_{t} 1/\sigma_{R_t}^2} }.
\end{equation}

This procedure explicitly performs the averaging over discrete time samples in the chosen pre-merger interval, making use of inverse-variance weighting. %It enables robust extraction and statistical analysis of circular polarization states and provides stringent tests of parity symmetry in gravitatio\-nal wave physics.

\section{Results}\label{sec:Results}

We analyzed the LVK parameter-estimation samples from the public Gravitational-Wave Open Science Center\cite{GWOSC} using our custom Python pipelines \cite{GWpolarizations-code}.

Out of all the events during which all 3 detectors were operational, we selected only those that had a narrow (``spot'') sky location\footnote{As a selection criterion, we included only those events for which the 50\% credible region of the sky localization formed a small, approximately circular spot with an angular diameter of a few degrees.}, because only in this case all 3 detectors captured a signal.

\begin{figure}
    \centering
    \includegraphics[width=0.48\textwidth]{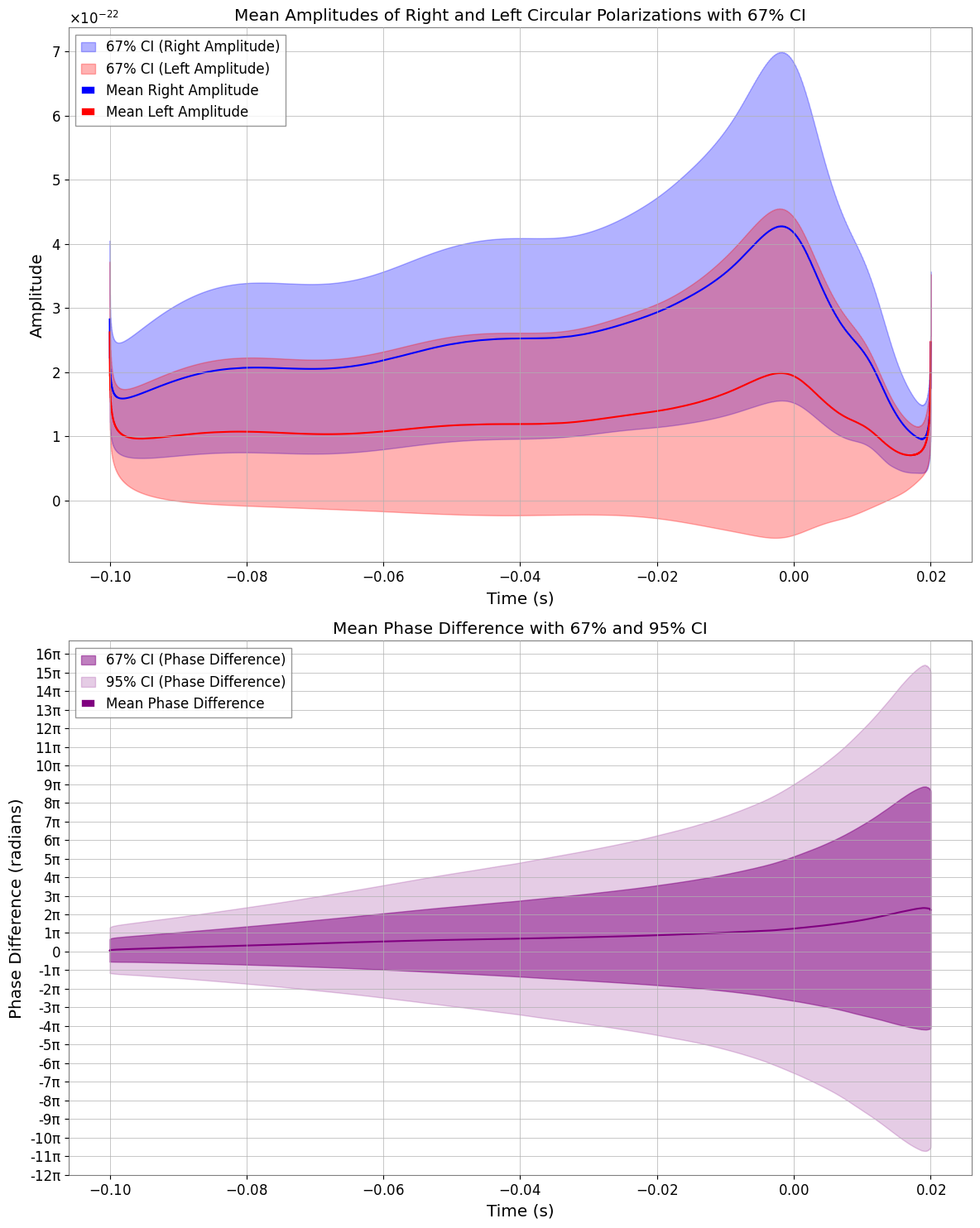} % Замените filename.png на имя файла
    \caption{GW190413 - more likely right than left (r), mean of one polarization is inside range of other polarization, but right amplitude dominates. Consistency check: phase difference between left and right polarization is constant within $1 \sigma$. }
    \label{190413}
\end{figure}

\begin{figure}
    \centering
    \includegraphics[width=0.48\textwidth]{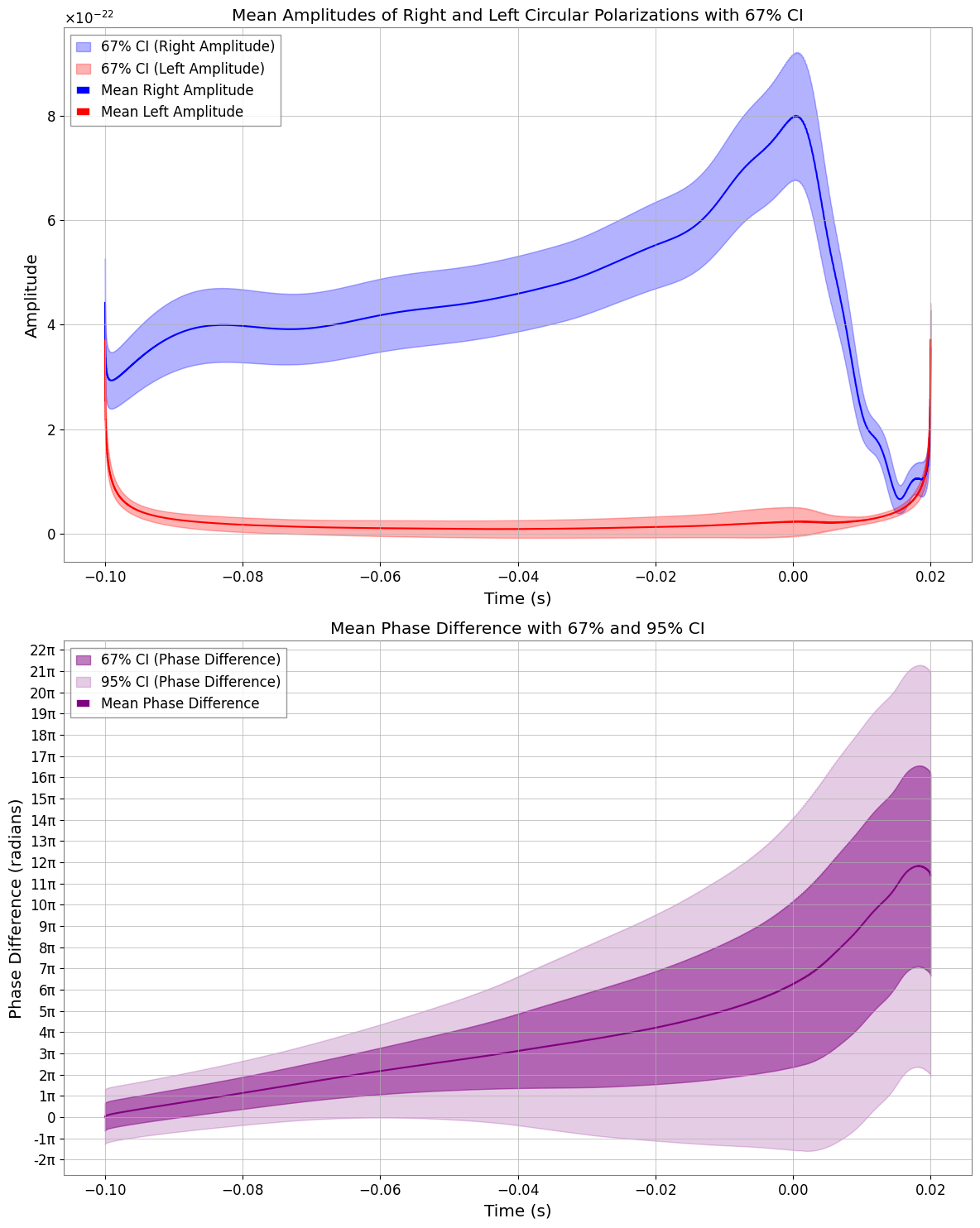} % Замените filename.png на имя файла
    \caption{GW200208 - right (R) - means of one polarization don't intersect the range of other polarization, right amplitude dominates. Consistency check: phase difference between left and right polarization is constant within $2 \sigma$.} 
    \label{200208}
\end{figure}

\begin{figure}
    \centering
    \includegraphics[width=0.48\textwidth]{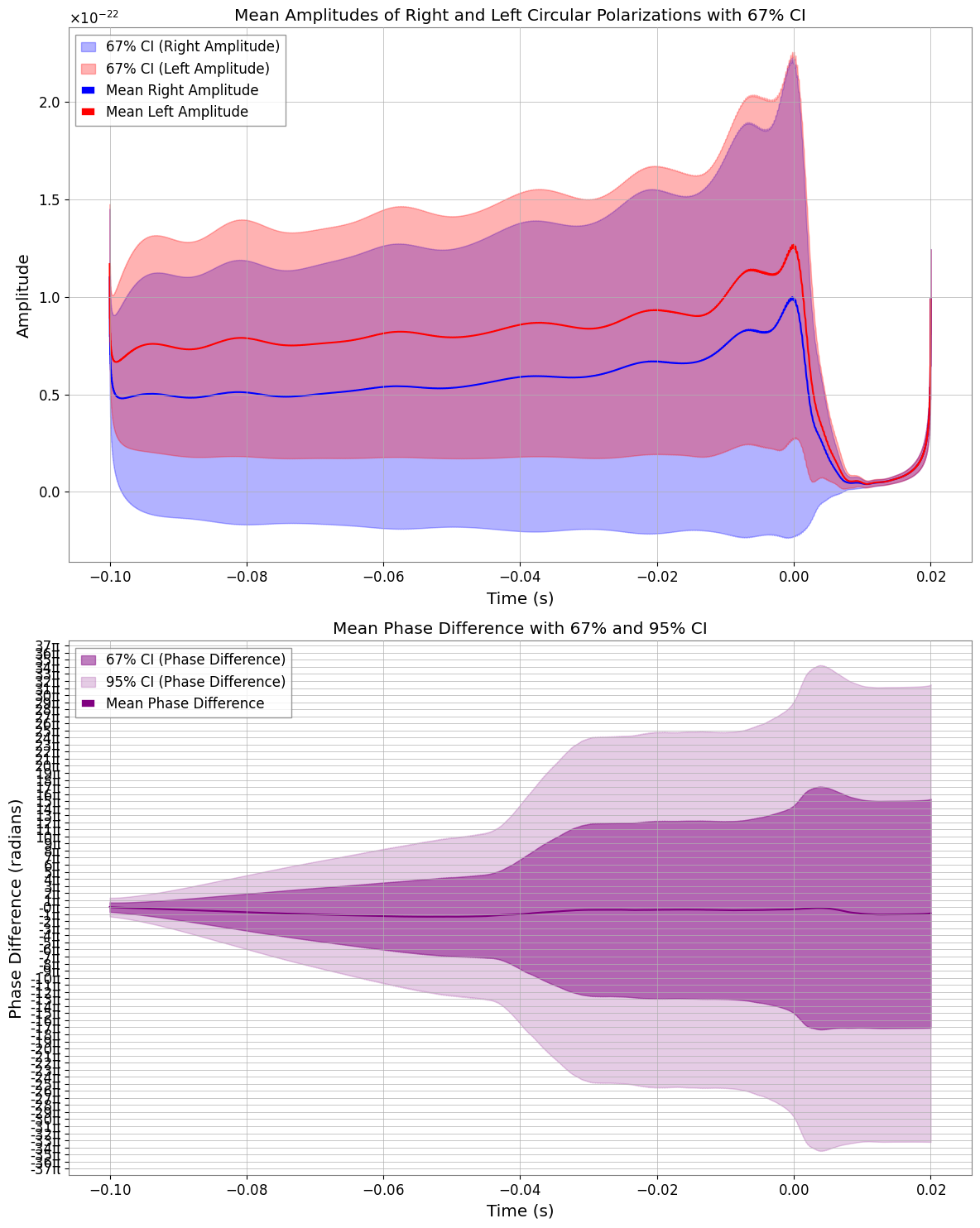} % Замените filename.png на имя файла
    \caption{GW200210 - more likely left than right (l), mean of one polarization is inside range of other polarization, but  left amplitude dominates. Consistency check: phase difference between left and right polarization is constant within $1 \sigma$.} 
    \label{200210}
\end{figure}

\begin{figure}
    \centering
    \includegraphics[width=0.48\textwidth]{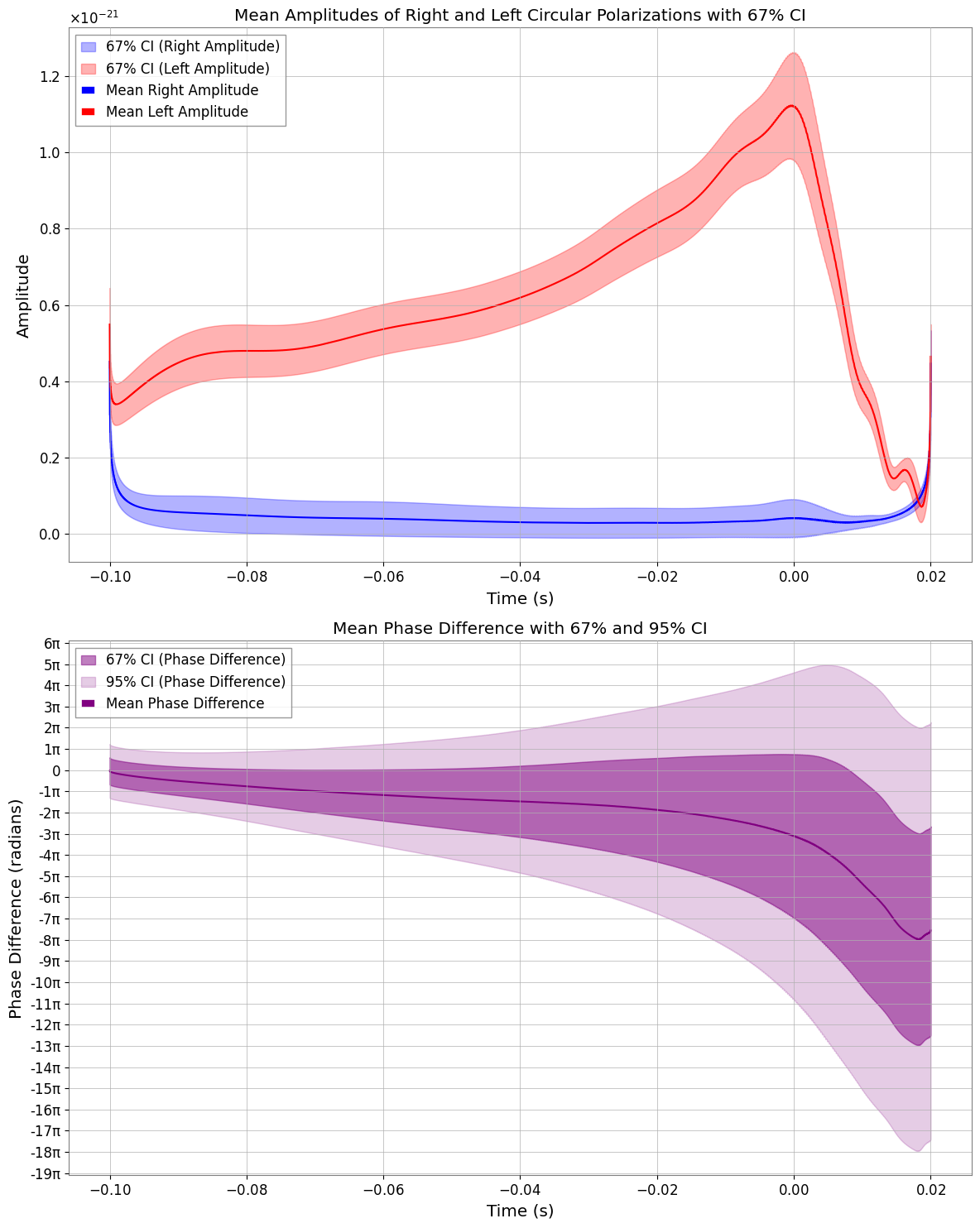} % Замените filename.png на имя файла
    \caption{GW200224 - left (L) - means of one polarization don't intersect the range of other polarization, left amplitude dominates. Consistency check: phase difference between left and right polarization is constant within $1 \sigma$.} 
    \label{200224}
\end{figure}

\noindent Figures~ \ref{190413}-\ref{200224}
illustrate representative events spanning our four classification labels: ``R/L'' for clear right/left dominance and ``r/l'' for ``more-likely'' cases where the posterior-mean amplitude of one helicity lies within the central credible range of the other. In each panel, we plot the reconstructed time-domain amplitudes \(A_{\MyCircleArrowRight}(t)\) and \(A_{\MyCircleArrowLeft}(t)\) (analytic-signal envelopes) over the pre-merger window used for averaging (see Sec. \ref{subsec:calcpol}), with shaded bands denoting 16--84\% credible intervals across posterior samples. The vertical band marks the interval contributing to the inverse-variance weighted time average \(\bar{R}\) in Eq. (\ref{averageratio}). ``R/L'' labels correspond to\\ \(\bar{A}_{\MyCircleArrowRight}/\bar{A}_{\MyCircleArrowLeft}>1\) or \(<1\), respectively; the lowercase ``r/l'' indicates overlap of the credible bands despite the mean ratio favoring one helicity.

For consistency across all events, we used the model \texttt{IMRPhenomXPHM} \cite{Pratten:2020ceb} as the default waveform model in our analysis.
Unfortunately, for most events in the O3a observing run, posterior samples based on \texttt{SEOBNRv4PHM} \cite{Ossokine:2020kjp} were unavailable, making that model inapplicable for the majority of our sample; however for 21 events we did also performed the polarization decomposition using the \texttt{SEOBNRv4PHM} approximant.
The resulting amplitude ratios and handedness classifications were consistent between the two models, although small differences were observed; these discrepancies between models can serve as estimators for systematic uncertainties, which we will not explore further at present.
These effects may influence the detailed amplitude ratios, particularly in cases near unity or for marginal detections, and should be considered in future high-precision polarization studies.
As a result, represented in table \ref{tab:gw_data}, we have 7 right and 8 left, 1 more likely right and 1 more likely left events with good sky location which are included into the statistics. Events without numerical value are not included. The averaged amplitude ratio logarithm is $\left<\log_{10}(A_R/A_L)\right> = -0.121 \pm 0.003.$
We see that we have too few well-localized events which we can use in the statistics. But we hope that in O4b\footnote{in O4a Virgo was not operational}, O4c and O5 this number will have increased.

%\clearpage

\section{Bayesian Estimation of Polarization Asymmetry}\label{sec:Bayes}

We consider the problem of inferring the true probability \( p_R \in [0,1] \) of observing a right-handed gravitational-wave polarization event, given a dataset consisting of \( N = 15 \) independent observations, among which \( k = 7 \) were right-handed and \( N - k = 8 \) were left-handed.

%\clearpage

\paragraph{Likelihood Function.}
Assuming that each detection is an independent Bernoulli trial with success probability \( p_R \), the likelihood function for observing \( k \) right-handed events out of \( N \) total events follows a binomial distribution:
\begin{equation}
    \mathcal{L}(p_R) \equiv P(\text{data} \mid p_R) = \binom{N}{k} p_R^k (1-p_R)^{N-k},
\end{equation}
where \( \binom{N}{k} \) is the binomial coefficient.

\paragraph{Prior Distribution.}
We adopt a noninformative (uniform) prior over \( p_R \):
\begin{equation}
    P(p_R) = 
    \begin{cases}
      1, & 0 \leq p_R \leq 1,\\
      0, & \text{otherwise}.
    \end{cases}
\end{equation}

\paragraph{Posterior Distribution.}
By Bayes' theorem, the posterior distribution for \( p_R \) is proportional to the product of the prior and the likelihood:
\begin{equation}
    P(p_R \mid \text{data}) = \frac{P(\text{data} \mid p_R) P(p_R)}{P(\text{data})},
\end{equation}
where \( P(\text{data}) \) is the evidence (normalization constant):
\begin{equation}
    P(\text{data}) = \int_0^1 P(\text{data} \mid p_R) P(p_R) \, dp_R.
\end{equation}

Given the uniform prior, the posterior simplifies to:
\begin{equation}
    P(p_R \mid \text{data}) = \frac{p_R^k (1-p_R)^{N-k}}{B(k+1, N-k+1)},
\end{equation}
where \( B(\alpha, \beta) \) is the Beta function:
\begin{equation}
    B(\alpha, \beta) = \int_0^1 p^{\alpha-1} (1-p)^{\beta-1} \, dp,
\end{equation}
and in our case \(\alpha = k+1 = 8\), \(\beta = N-k+1 = 9\).

Thus, the posterior distribution is the Beta distribution:
\begin{equation}
    p_R \mid \text{data} \sim \text{Beta}(8,9).
\end{equation}

\paragraph{Credible Intervals.}
To summarize the posterior distribution, we compute central credible intervals corresponding to confidence levels associated with \(1\sigma\), \(2\sigma\), and \(3\sigma\) under the assumption of approximate normality:

\begin{itemize}
    \item The 68.27\% ($\sim 1\sigma$) credible interval, the region between the 15.865th and 84.135th percentiles of the posterior distribution, is
    %:
    %\begin{equation}
        $p_R \in (0.350, 0.592)$.
    %\end{equation}
    \item The 95.45\% ($\sim 2\sigma$) credible interval, between the 2.275th and 97.725th percentiles, is %:
    %\begin{equation}
        $p_R \in (0.243, 0.706)$.
    %\end{equation}
    \item The 99.73\% ($\sim 3\sigma$) credible interval, between the 0.135th and 99.865th percentiles, is %:
%    \begin{equation}
        $p_R \in (0.155, 0.802)$.
    %\end{equation}
\end{itemize}

These intervals indicate the ranges within which the true right-handed polarization probability \( p_R \) lies with the corresponding posterior probabilities.

\paragraph{Tail Probabilities.}
Additionally, we compute posterior probabilities that \( p_R \) exceeds specific thresholds:
\begin{align}
    P(p_R \geq 0.6) &= 1 - F(0.6) \approx 14.2\%, \\
    P(p_R \geq 0.9) &= 1 - F(0.9) \approx 5.9 \times 10^{-6}, \\
    P(p_R \geq 0.99) &= 1 - F(0.99) \approx 1.1 \times 10^{-14},
\end{align}
where \( F(p) \) is the cumulative distribution function (CDF) of the Beta(8,9) distribution.

These results imply that the observed data strongly disfavor hypotheses with an extremely high intrinsic right-handed polarization probability (\( p_R \gtrsim 0.9 \)).

\paragraph{Posterior Plot.}
Figure~\ref{fig:posterior} shows the posterior probability density function of \( p_R \), with shaded regions corresponding to the \(1\sigma\), \(2\sigma\), and \(3\sigma\) credible intervals, and vertical dashed lines marking their boundaries.

\begin{table}[h]
    \centering
   \begin{tabular}{ccccc}
        \hline
         GW &
         Amplitude &
         Pol.  & Location&Obs.\\
        Event &
        \textbf{\( A_R/A_L \)} &
        P(S) & \!\!\!\!\! Skymap& \!\!\!\!\! Run\\
        \hline
     \!\!\!\!\!GW200322 &  & r(R) & $\bigcirc$&O3b \\
        \!\!\!\!\!GW200316 & $18.29 \pm 0.59 $ & R(R) & $\cdot$&O3b \\
        \!\!\!\!\!GW200311 & $ (33.33 \pm 1.11)^{-1} $ & L(L) & $\cdot$&O3b\\
        \!\!\!\!\!GW200308 &  & l(L) & $\bigcirc$&O3b \\
        \!\!\!\!\!GW200224 & $ (20.41 \pm 0.42)^{-1} $ & L(L) & $\cdot$&O3b\\
        \!\!\!\!\!GW200220\_06 &  & l(L) & $\int$&O3b\\
        \!\!\!\!\!GW200219 &  & L(L) & $\int$&O3b\\
        \!\!\!\!\!GW200216 &  & r(l) & $\int$&O3b\\
        \!\!\!\!\!GW200210 &  & l(r) & $\int$&O3b\\
        \!\!\!\!\!GW200209 &  & r(l) & $\int$&O3b\\
        \!\!\!\!\!GW200208\_22 &  & L(l) & $\int$&O3b\\
        \!\!\!\!\!GW200208\_13 & $21.95 \pm 0.45$ & R(R) & $\cdot$&O3b\\
        \!\!\!\!\!GW200202 & $ 40.14 \pm 1.27 $ & R(R) & $\cdot$&O3b\\
        \!\!\!\!\!GW200129 & $(55.56 \pm 3.09)^{-1}$ & L(L) & $\cdot$&O3b\\
        \!\!\!\!\!GW200115 & $(37.04 \pm 1.37)^{-1}$  & L &  $\cdot$&O3b\\
        \!\!\!\!\!GW191230 &  & R(r) & $\int$&O3b\\
        \!\!\!\!\!GW191219 &  &  (r) & $\int$&O3b\\
        \!\!\!\!\!GW191215 &  & l(l) & $\int$&O3b\\
        \!\!\!\!\!GW191127 &  & l(L) & $\int$&O3b\\
        \!\!\!\!\!GW191113 &  & r(r) & $\int$&O3b\\
        \!\!\!\!\!GW191105 &  & L(L) & $\int$&O3b\\
        \!\!\!\!\!GW190930 &  & L(L) & $\int$&O3a\\
        \!\!\!\!\!GW190929 &  & L  & $\int$&O3a\\
        \!\!\!\!\!GW190926 &  & r  & $\int$&O3a\\
        \!\!\!\!\!GW190924 &  & L  & $\int$&O3a\\
        \!\!\!\!\!GW190917 &  & l  & $\int$&O3a\\
        \!\!\!\!\!GW190916 &  & l  & $\int$&O3a\\
        \!\!\!\!\!GW190915 &  & l  & $\int$&O3a\\
        \!\!\!\!\!GW190828\_0655 &  & r  & $\int$&O3a\\
        \!\!\!\!\!GW190828\_0634 &  & R &  $\int$&O3a\\
        \!\!\!\!\!GW190814 &  $(15.15 \pm 0.46)^{-1}$  & L  & $\cdot$&O3a\\
        \!\!\!\!\!GW190803 &  & l  & $\int$&O3a\\
        \!\!\!\!\!GW190728 &  $(1.43 \pm 0.02)^{-1}$  & l  & $\int$&O3a\\
        \!\!\!\!\!GW190727 &  $1.641 \pm 0.038$  & r  & $\int$&O3a\\
        \!\!\!\!\!GW190725 &  & L  & $\int$&O3a\\
        \!\!\!\!\!GW190720 &  $39.86 \pm 1.15$  & R  & $\cdot$&O3a\\
        \!\!\!\!\!GW190706 &  & l  & $\int$&O3a\\
        \!\!\!\!\!GW190701 &  $(50.00 \pm 2.50)^{-1}$  & L  & $\cdot$&O3a\\
        \!\!\!\!\!GW190602 &  & r &  $\int$&O3a\\
        \!\!\!\!\!GW190521\_03 &  & l  & $\int$&O3a\\
        \!\!\!\!\!GW190519 &  & r  & $\int$&O3a\\
        \!\!\!\!\!GW190517 &  & l  & $\int$&O3a\\
        \!\!\!\!\!GW190513 &  & L  & $\int$&O3a\\
        \!\!\!\!\!GW190512 &  & l  & $\int$&O3a\\
        \!\!\!\!\!GW190503 & \( 13.79 \pm 0.40 \) & R &  $\cdot$&O3a\\
        \!\!\!\!\!GW190413\_13 &  & r &  $\int$&O3a\\
        \!\!\!\!\!GW190412 & \( (38.46 \pm 1.48)^{-1} \) & L  & $\cdot$&O3a\\
        \!\!\!\!\!GW190408 &  & L  & $\int$&O3a\\
        \!\!\!\!\!GW170818 &  $22.01 \pm 0.46$  & R & $\cdot$&O2\\
        \!\!\!\!\!GW170814 &  $(16.67 \pm 0.56)^{-1}$  & L  & $\cdot$&O2\\
        \!\!\!\!\!GW170809 &  $26.99 \pm 0.43$  & R & $\cdot$&O2\\
        \!\!\!\!\!GW170729 &  & L  & $\int$&O2\\
        \hline
    \end{tabular}
    \caption{The analyzed events,
    % are
    given in reverse chronological order.%, starting from the latest at the top.
    Polarizations were calculated using two waveform template models: P (Phenom, specifically, IMRPhenomXPHM) and S (SEOB, specifically, SEOBNRv4PHM). The skymap types are~ $\bigcirc$  - effectively all sky,~ $\int$ - banana-shaped strip,~ $\cdot$ - narrow spot, almost a dot.}
    \label{tab:gw_data}
\end{table}

\subsection{ Consistency of the observed result with binomial expectations}

We examine the consistency of the observed number of right-handed events, \( k = 7 \) out of \( N = 15 \), with different hypotheses for the true polarization probability \( p_R \), under a binomial likelihood model.

For a fixed value of \( p_R \), the probability of observing exactly \( k \) right-handed events is given by the binomial probability mass function:
\begin{equation}
    P(k \mid p_R) = \binom{N}{k} p_R^k (1 - p_R)^{N - k}.
\end{equation}

To assess which values of \( p_R \) are consistent with the observed data at the 1$\sigma$ and 2$\sigma$ levels, we determine the values of \( p_R \) for which \( k = 7 \) lies within the central credible interval of the binomial distribution:
\begin{equation}
    P(k_{\text{low}} \leq k \leq k_{\text{high}} \mid p_R) = \gamma,
\end{equation}

\clearpage

\!\!\!\!\!\!\!\!\!\!

where \( \gamma = 0.6827 \) (68.27\%) for 1$\sigma$ and \( \gamma = 0.9545 \) (95.45\%) for 2$\sigma$.

Using this method, we find that:

\begin{itemize}
    \item At the 1$\sigma$ level (central 68.27\% binomial interval), the observed result \( k = 7 \) is consistent with values of \( p_R \) in the range:
    \[
    p_R \in (0.314,\ 0.624).
    \]
    \item At the 2$\sigma$ level (central 95.45\% binomial interval), the result is consistent with:
    \[
    p_R \in (0.209,\ 0.738).
    \]
\end{itemize}

These intervals indicate that strongly asymmetric polarization probabilities (e.g., \( p_R \lesssim 0.2 \) or \( p_R \gtrsim 0.75 \)) are inconsistent with the observed result at the 2$\sigma$ level.

\subsection{Forecast for future observations: maximum deviation consistent with symmetry}

Assume that in a future observation run, a total of \( N = 100 \) well-defined gravitational-wave events are detected with confidently measured polarizations. Under the null hypothesis that the true right-handed polarization probability is symmetric, i.e., \( p_R = 0.5 \), we can predict the expected statistical spread in the observed number of right-polarized events using the binomial distribution.

For large \( N \), the binomial distribution can be approximated by a normal distribution with mean and standard deviation:
\begin{align}
    \mu &= N p_R = 50, \\
    \sigma &= \sqrt{N p_R (1 - p_R)} = \sqrt{100 \cdot 0.5 \cdot 0.5} = 5.
\end{align}

The 68.27\% (1$\sigma$) and 95.45\% (2$\sigma$) confidence intervals for the number of right-handed events are therefore:
\begin{align}
    \text{1$\sigma$ interval:} & \quad k \in [\mu - \sigma, \mu + \sigma] = [45,\ 55], \\
    \text{2$\sigma$ interval:} & \quad k \in [\mu - 2\sigma, \mu + 2\sigma] = [40,\ 60].
\end{align}

\begin{figure}[h]
    \centering
    \includegraphics[width=1\linewidth]{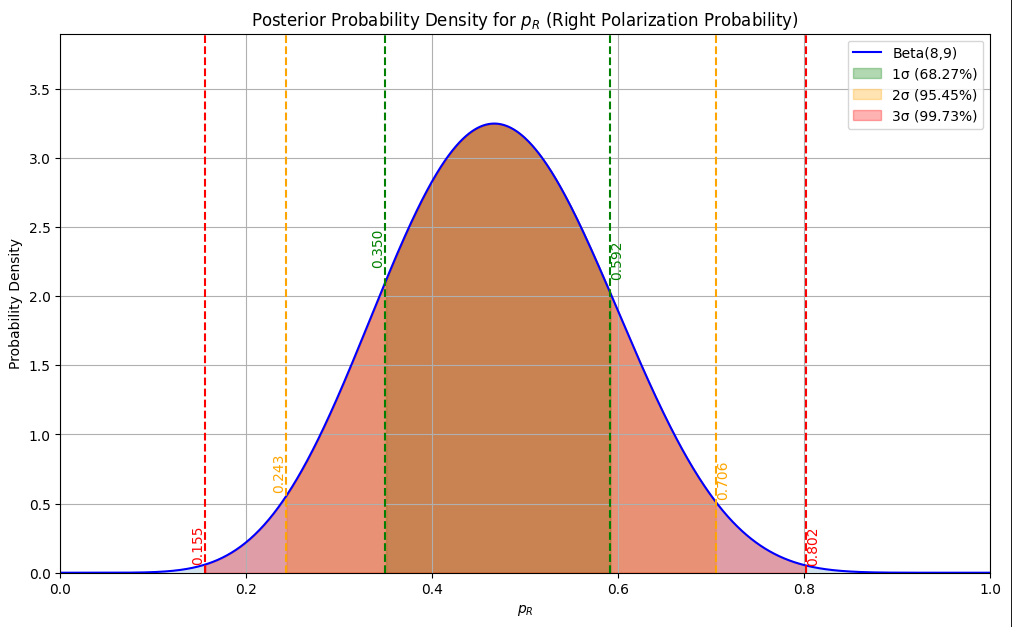}
    \caption{Posterior probability density \(P(p_R \mid \text{data})\) for the right-handed polarization probability, based on 7 right-handed and 8 left-handed gravitational-wave events. Shaded regions and dashed lines indicate \(1\sigma\) (68.27\%), \(2\sigma\) (95.45\%), and \(3\sigma\) (99.73\%) credible intervals.}
    \label{fig:posterior}
\end{figure}

\begin{figure}[h]
    \centering
    \includegraphics[width=1\linewidth]{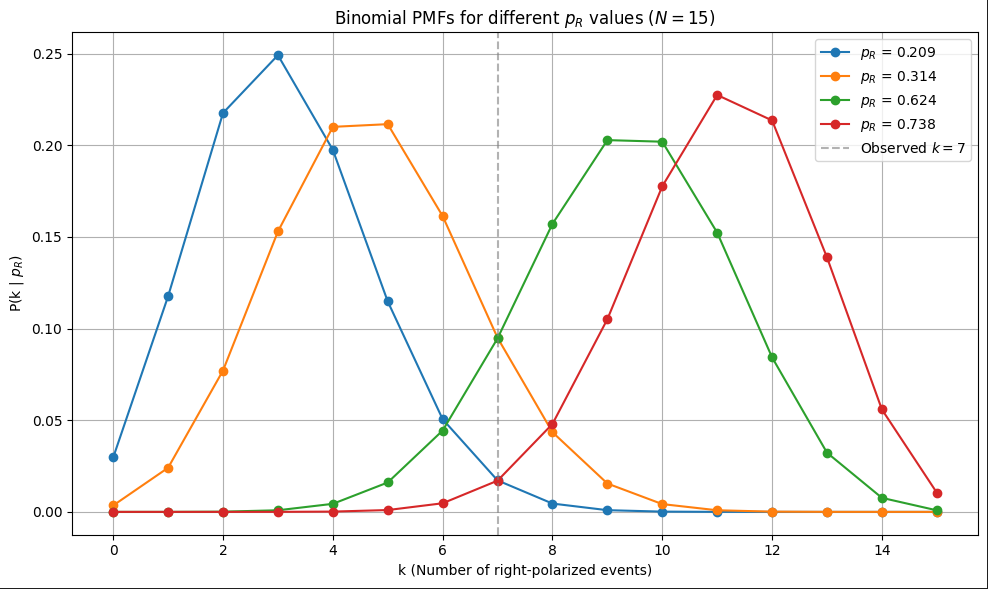}
    \caption{Binomial probability mass functions \( P(k \mid p_R) \) for several values of the true right-handed polarization probability \( p_R \), with \( N = 15 \) total events. Each curve shows the likelihood of observing \( k \) right-handed events under a specific \( p_R \). The vertical dashed line marks the observed value \( k = 7 \). The values \( p_R = 0.314 \) and \( p_R = 0.624 \) correspond to the edges of the 68.27\% credible interval (1$\sigma$), while \( p_R = 0.209 \) and \( p_R = 0.738 \) correspond to the 95.45\% interval (2$\sigma$). The figure illustrates how different hypotheses for \( p_R \) place the observed result within or outside their typical ranges.}
    \label{fig:posterior}
\end{figure}

In terms of observed right-handed fractions \( \hat{p}_R = k / N \), these intervals correspond to:
\begin{align}
    \text{1$\sigma$:} & \quad \hat{p}_R \in [0.450,\ 0.550], \\
    \text{2$\sigma$:} & \quad \hat{p}_R \in [0.400,\ 0.600].
\end{align}

Therefore, if the true value is \( p_R = 0.5 \), then observing a polarization ratio \( \hat{p}_R \) within the range \([0.450,\ 0.550]\) would be consistent at the 1$\sigma$ level, and within \([0.400,\ 0.600]\) at the 2$\sigma$ level. Observed deviations beyond these bounds would suggest a statistically significant departure from perfect left–right symmetry.
\subsection{Future Prospects: Probing Symmetry Breaking with More Events}

While our current analysis indicates no statistically significant deviation from circular polarization symmetry, the limited number of well-localized binary black hole (BBH) events severely restricts our sensitivity to potential asymmetries.

To quantify future prospects, we consider a simplified model in which the underlying probability to observe a right-handed gravitational-wave polarization is \( p_R = 7/15 \approx 0.4667 \), matching the observed ratio, and the left-handed probability is \( p_L = 1 - p_R = 8/15 \approx 0.5333 \). We ask: how many events must be observed before the hypothesis of symmetry \( (p_R = 0.5) \) becomes statistically inconsistent with the true asymmetry at various confidence levels?

Since the number of detections will be large, the binomial distribution for the number of right-handed events \( k \) among \( N \) total observations can be approximated by a normal distribution with mean and variance:
\begin{equation}
    \mu = N p_R, \quad \sigma^2 = N p_R (1 - p_R).
\end{equation}
We then compute the standardized distance \( z \) between the symmetric prediction \( k = N/2 \) and the mean of the asymmetric model:
\begin{equation}
    z = \frac{N/2 - N p_R}{\sqrt{N p_R (1 - p_R)}} = \sqrt{N} \times \frac{(1/2 - p_R)}{\sqrt{p_R(1 - p_R)}}.
\end{equation}
Substituting \( p_R = 7/15 \), we obtain:
\begin{equation}
    \frac{1/2 - p_R}{\sqrt{p_R (1 - p_R)}} \approx 0.0668.
\end{equation}
Thus, the number of events required to exclude \( p_R = 0.5 \) at a given significance level \( z_\alpha \) is:
\begin{equation}
    N \geq \left(\frac{z_\alpha}{0.0668}\right)^2.
\end{equation}
Using standard values for \( z_\alpha \), we find:

\begin{itemize}
    \item For \( 1\sigma \) (68.27\% confidence, \( z \approx 1 \)): \quad \( N \geq 224 \),
    \item For \( 2\sigma \) (95.45\% confidence, \( z \approx 2 \)): \quad \( N \geq 894 \),
    \item For \( 3\sigma \) (99.73\% confidence, \( z \approx 3 \)): \quad \( N \geq 2017 \).
\end{itemize}

Therefore, if the true underlying asymmetry is mild (e.g., \( p_R = 7/15 \)), then several hundred to a few thousand well-localized BBH events are required to statistically distinguish it from perfect symmetry at the 1–3$\sigma$ levels.

Considering the projected detection rates for the recent and upcoming observing runs (O4b, O4c, O5 and beyond) and the expansion of the global detector network (with KAGRA increasing in sensitivity and LIGO–India expected to join), accumulating one to several hundred three‑detector events over the next several years is realistic \cite{LIGO-Observing,MiningO4Alerts}. The detection of polarization asymmetry would provide valuable insight into potential parity violation, birefringence, or extensions beyond General Relativity.

\section{Conclusions}\label{sec:Concl}

In this work, we have investigated the possibility of parity symmetry breaking in gravitational-wave signals from binary black hole (BBH) mergers, focusing on potential asymmetries between right-handed and left-handed circular polarizations. Motivated by the fundamental importance of parity violation in particle physics and by theoretical extensions of General Relativity, we reanalyzed publicly available gravitational-wave events from the LIGO-Virgo-KAGRA \newline collaboration.

We reviewed the key role of three-detector networks in enabling precise sky localization and polarization separation, both of which are critical for measuring the circular polarization content of gravitational-wave signals. Using a sample of well-localized events, we performed a polarization decomposition and statistically analyzed the distribution of handedness among detected BBH mergers.

Our current results, based on a limited number of events, show no statistically significant deviation from the expected parity symmetry. The slight observed imbalance between left- and right-handed polarizations (8 left versus 7 right events) is fully consistent with random fluctuations at small sample sizes. Consequently, no definitive conclusion regarding symmetry breaking can yet be drawn.

However, using a Bayesian framework with a uniform prior, we inferred that the probability \( p_R \) of observing a right-handed polarization follows a \(\text{Beta}(8,9)\) posterior distribution. The resulting credible intervals show that \( p_R \) most likely lies between 0.350 and 0.592 (68.27\% credibility), between 0.243 and 0.706 at 95.45\% credibility (\(2\sigma\)), and between 0.155 and 0.802 at 99.73\% credibility (\(3\sigma\)). Extremely high values of \( p_R \) (e.g., \( p_R \gtrsim 0.9 \)) are strongly disfavored by the observed data.

To assess the future prospects, we modeled a small intrinsic asymmetry between right- and left-handed polarizations (probabilities 8/15 and 7/15, respectively). Under this model, we find that approximately 2000 well-localized BBH events would be required for an equal number of right- and left-handed detections to be statistically inconsistent with the assumed asymmetry at the $3\sigma$ level (99.7\% confidence). Given expected improvements in detector sensitivity, increa\-sed event rates, and the expansion of the global detector network, achieving this number of high-quality events within the next few observing runs is realistic.

We plan to extend this polarization analysis to the events that will be detected in the upcoming observing runs (O4, O5 and beyond).

Therefore, the search for circular polarization asymmetries in gravitational waves remains an exciting and viable path for testing fundamental symmetries of gravity, with the potential to uncover new physics beyond General Relativity as future data accumulate.

%\begin{figure*}
%  \includegraphics[width=0.75\textwidth]{example.eps}
%\end{figure*}
%

\begin{acknowledgements}
EE has been supported in part by the Israeli Ministry of Aliyah and Integration,
by the “Program of Support of High Energy Physics” Grant by Israeli Council for Higher Education, and
by the Israel Science Fund (ISF) grant No. 1698/22.”.
The authors are grateful to Ido Ben-Dayan for discussions and to Viacheslav Prokopov 
 for help with the code.

In addition, This research has made use of data or software obtained from the Gravitational Wave Open Science Center (gwosc.org), a service of the LIGO Scientific Collaboration, the Virgo Collaboration, and KAGRA. This material is based upon work supported by NSF's LIGO Laboratory which is a major facility fully funded by the National Science Foundation, as well as the Science and Technology Facilities Council (STFC) of the United Kingdom, the Max-Planck-Society (MPS), and the State of Niedersachsen/Germany for support of the construction of Advanced LIGO and construction and operation of the GEO600 detector. Additional support for Advanced LIGO was provided by the Australian Research Council. Virgo is funded, through the European Gravitational Observatory (EGO), by the French Centre National de Recherche Scientifique (CNRS), the Italian Istituto Nazionale di Fisica Nucleare (INFN) and the Dutch Nikhef, with contributions by institutions from Belgium, Germany, Greece, Hungary, Ireland, Japan, Monaco, Poland, Portugal, Spain. KAGRA is supported by Ministry of Education, Culture, Sports, Science and Technology (MEXT), Japan Society for the Promotion of Science (JSPS) in Japan; National Research Foundation (NRF) and Ministry of Science and ICT (MSIT) in Korea; Academia Sinica (AS) and National Science and Technology Council (NSTC) in Taiwan.
%\\
Some of the results in this paper have been derived using the ``PESummary``
package.
%\\
This research has made use of Parallel Bilby \cite{bilby_paper,pbilby_paper}, a parallelised Bayesian inference Python package, and Dynesty \cite{dynesty_paper}, a nested sampler, to perform Bayesian parameter estimation and RIFT \cite{Pankow:2015cra,Lange:2018pyp}, Rapid parameter inference on gravitational wave sources.

\end{acknowledgements}

\bibliographystyle{spmpsci_unsrt}      % mathematics and physical sciences
\bibliography{references}   % name your BibTeX data base

\end{document}